\documentclass[10pt]{article}

\usepackage{graphicx}
\usepackage{amsmath}
\usepackage{url}
\usepackage{times}
\usepackage{helvetic}

\usepackage{amssymb}

\begin{document}
\pagestyle{empty}
\title{Characterizing correlations of flow oscillations at bottlenecks}

\author{Tobias Kretz, Marko W{\"o}lki, and Michael Schreckenberg\\
Physik von Transport und Verkehr,\\ Universit\"{a}t Duisburg-Essen,\\ 47057 Duisburg, Germany\\ 
\{kretz,woelki,schreckenberg\}@traffic.uni-duisburg.de\\
}

\maketitle

\begin{abstract}
``Oscillations" occur in quite different kinds of many-particle-systems when two groups
of particles with different directions of motion meet or intersect at a certain spot.
In this work a model of pedestrian motion is presented that is able to reproduce oscillations with different
characteristics. The Wald-Wolfowitz test and Gillis' correlated random walk are shown to include
observables that can be used to characterize different kinds of oscillations.
\end{abstract}

\section{Introduction}
The widely used word {\em oscillation} is defined for this article in a very 
general sense as the phenomenon that the direction of flow changes in
certain intervals when two groups of particles compete for the chance to move at a narrow
bottleneck. Oscillations in pedestrian dynamics have been dealt with before (see 
\cite{Helbing95} for example). The aim of this work however is to give a framework for 
an analytical and quantitative treatment of oscillations in observation, experiment and 
simulation, which has not been done before. There are three extreme types 
of oscillation: 1) the zipper principle: the flow direction changes
after each particle. 2) no oscillations: the flow direction only changes after one group has completely passed through the bottleneck.
3) uncorrelated oscillation: the statistics of the flow direction is the same as for coin-tossing experiments. While in the
first two scenarios the whole process is completely determined as soon as the first particle has passed the bottleneck, there
is absolutely no influence from one {\em event} to the next in the third scenario. Between those extremes there is a
continuous spectrum of correlated oscillations with an influence from one event to the other but no full determination.
The outline of the paper is as follows: First we present the parts of our model of pedestrian motion that are crucial for oscillations.
We then describe the test scenario in detail. The third step is the presentation of the observables that characterize a discrete oscillation.
Finally we present the simulation results for different sets of parameters.

\section{The model}
A model of pedestrian motion is presented now that expands the model of \cite{Burstedde01,Nishinari04} for velocities $v>1$ and adds some 
more features. Here we will only 
present the elements of the model that are involved in the phenomenon of oscillations. The full model in all details will
be presented elsewhere \cite{Kretz:phd}. The model is discrete in space ({\em cells}) as well as in time ({\em rounds}). Space is 
discretized into a regular lattice with cells with border length 40 cm.
Each cell may be occupied by at most one {\em agent} as we will call the representation of a real person in the 
model. At the beginning of the simulation the distance of each cell to each exit is calculated and stored in the integer valued {\em static
floor field} $S_{xy}^e$. So $S_{xy}^e$ is the distance of cell $(x,y)$ to exit $e$. 
Additionally to the static floor field there is an also integer valued
{\em dynamic floor field} $\vec{D}_{xy}=(D_{xy}^x,D_{xy}^y)$ which in our model is a vector field. Whenever an agent leaves 
cell $(x,y)$ and moves to cell $(x+i,y+j)$ the vector $(i,j)$ is added to the dynamic floor field at the position $(x,y)$. In between the 
rounds of the agents' motion the dynamic floor field is subject to diffusion and decay with the two components diffusing and decaying independently: 
each entry (positive or negative) in a component decays with probability $\delta$ and if it does not decay, it diffuses with 
probability $\alpha$ to one of the neighboring cells.\\
The agents can have different maximal velocities. This is implemented by the number of cells an agent is allowed to move into any direction
during one round. As one round is interpreted as one second, typical velocities are two to five cells per round. See 
\cite{PED05:Kretz} for details. So for each agent $a$ there is a set $R^a$ of cells, which are unoccupied and visible (not hidden by a wall or 
another agent) and which can be reached during the next round. At the beginning of each round each agent $a$ randomly chooses a
{\em destination cell} $(x_d^a,y_d^a)$ out of $R^a$. The probability for a cell $(x,y)$ within $R^a$ to be chosen as 
destination cell by an agent currently positioned at $(x_0^a,y_0^a)$ is:
\begin{equation}
p_{x,y}^a \propto \exp\Big(k_S^a S_{xy}^{e=f(a,x_0^a,y_0^a,x_e,y_e)} + k_D^a\big((x-x_0^a)D^x_{x_0^a,y_0^a}+(y-y_0^a)D^y_{x_0^a,y_0^a}\big)\Big)
\end{equation}
Compared to the model variants in \cite{Kirchner04} where desired trajectories are calculated in advance, by applying the
same update rules $v_{max}$ times and then different possible ways to use that trajectory to reach the final cell are investigated, 
here only a desired destination cell is calculated and the actual movement process is not implemented as planning process but as
``real" motion. Note especially the equality of all cells of $R^a$, where in principal the relative position of a cell to an agent
does not matter.\par
The dynamic floor field therefore influences the motion via the scalar product with a possible direction of motion. $S_{xy}^{e=f(a,x_0^a,y_0^a,x_e,y_e)}$ means that in the presence of more than one exit the choice of an exit is a function of the
number (group affiliation) and the position of the agent as well as the position of the exit. Here all agents are only allowed
to use the exit on the other side of the bottleneck compared to their starting position.\par
After all agents have chosen their destination cell in parallel, the agents start to move sequentially. It is important to mention that
on their way to their destination cell they execute {\em single steps} within the Moore neighborhood \cite{Wolfram:taaca} 
which implies that for $v=1$ one step within
the Moore neighborhood is possible. The sequence of all single steps of all
agents is fixed randomly at the beginning of the motion part of a round. Compared to the model treated in 
\cite{Woelki06} there is an additional 
shuffle as an agent does not necessarily execute all his steps at once. Whenever an agent has stepped on a cell and moved on, the cell
remains {\em blocked} for all other agents until the rest of the round. This represents the speed-dependent occupation 
of space during motion \cite{Kluepfel:phd,ped05:Seyfried,Kirchner04}, which is essential for a realistic fundamental diagram. 
Within a Moore neighborhood an agent always chooses the cell which lies closest to
his destination cell and which has not been blocked before by another agent. If there is no unblocked cell left which is closer 
to the destination cell than the current cell, the round ends for that particular agent. Therefore an agent can fail to reach
his destination cell. If two or more agents choose the same cell as destination cell or need the same cell as part of their
path to their destination cell in the calculations for this work simply the order of movement 
decides about the winner of such a ``conflict".\par

\section{The scenario}
\begin{figure}[htbp]
\begin{center}
\includegraphics[width=203pt]{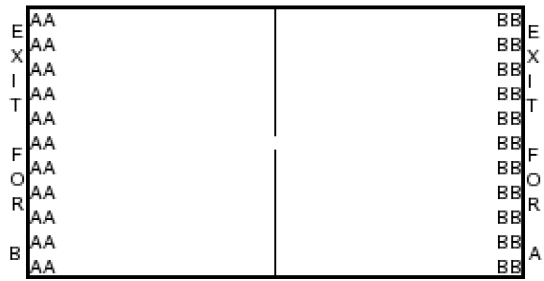}
\caption{Initial positions of the agents in the scenario (schematic view).}
\label{fig:startosc}
\end{center}
\end{figure}
At the beginning of the simulation two groups ($A$ and $B$) with $N_A=N_B=202$ ($N=N_A+N_B$) agents each are placed in two 
rows as shown in figure (\ref{fig:startosc}). They have to pass the bottleneck with a width
of one cell in the middle and proceed to the exit on the other side. An agent is counted as having passed as soon as he
leaves the bottleneck cell in the direction of his exit. The possibility that he moves back into the bottleneck is given but neglected.
Only the first $n=100$ agents passing the bottleneck are taken into account to
make sure that there are always two almost equally sized groups competing for passage. Of these $n=100$ agents the number 
of agents of group $A$ is called $n_A$ and of group $B$ $n_B$. $n$ can not be chosen arbitrarily large, as at some size 
the crowds in front of the bottleneck become that dense that no passing at all is possible. 
\par
There are no new agents entering the scenario.

\section{Observables}
\subsection{The Wald-Wolfowitz test}
% The Wald-Wolfowitz test (also called {\em ``runs test"}) 
The Wald-Wolfowitz test can be used to check for correlations in time series whenever the statistic is dichotomous or can be dichotomized, 
i.e. when two kinds of events or data are present. The probabilities for the events to happen do not need to be equal.
In our case those two kinds of events are ``A member of group $A$ passes the bottleneck" and ``A member of group $B$ passes the bottleneck". The
Wald-Wolfowitz test makes a statement about the expectation value and the variance of the number of runs (and other observables) if 
there are no correlations between the different events. In the terminology of the test a {\em run} is a series of events 
of one kind not interrupted by an event of the other kind.\\
The crucial quantities of the Wald-Wolfowitz test are:
\begin{itemize}
\item $\langle R(n_A,n_B) \rangle$ is - for given $n_A$ and $n_B$ - the expectation value of the number of runs for an uncorrelated series of - in our case - bottleneck passages. \vspace{-6pt}
\item $\sigma_R$ is the standard deviation of the number of runs.\vspace{-6pt}
\item $r$ is the number of runs in a certain simulation,\vspace{-6pt}
\item $z=\frac{r-\langle R(n_A,n_B) \rangle}{\sigma_R}$ is the standardized test variable.
\end{itemize}
For $n_A,n_B \gg 1$ (at least $n_A,n_B > 10$) the distribution of the number of runs becomes comparable to the normal distribution 
which implies that the null 
hypothesis {\em ``The events are uncorrelated"} can be rejected on an $\alpha$-level of significance if $|z|>q(1-\alpha/2)$, 
with $q(1-\alpha/2)$ as quantile of the standard 
normal distribution for the probability $1-\alpha/2$. For smaller $n_A,n_B$ one has to make use of tables to decide about the rejection
of the null hypothesis \cite{runstest:daniel,runstest:Bradley}. \\

For an uncorrelated process one has for given $n_A,n_B$
\begin{eqnarray}
\langle R(n_A,n_B) \rangle&=&2\frac{n_An_B}{n}+1=2\frac{n_A(n-n_A)}{n}+1 \label{eq:R}\\
%\overline{\langle R\rangle }&=&\Sum_{n_A=0}P(n_A)\langle R(n_A,n_B) \rangle\\
\sigma_R^2&=&\frac{2n_An_B(2n_An_B-n)}{n^2(n-1)}=\frac{(\langle R \rangle-1)(\langle R \rangle-2)}{n-1}
\end{eqnarray}

The Wald-Wolfowitz test is independent of underlying distributions; therefore it does not make use of 
deviations in the distribution of $n_A$ in subsequently repeated simulations of a scenario. This additional information
can be made use of by mapping the process onto a correlated random walk.

\subsection{Correlated random walk} 
In correlated random walk models \cite{Gillis55,Renshaw92} the probabilities for the direction of the next step depend on 
the direction of the last step. 
The ``correlated random walker" keeps his direction of motion with probability $p$ and changes it with probability $1-p$. In the
first step the direction is chosen with equal probability.
In the thermodynamic limit ($n \rightarrow \infty$) 
one then gets a normal distribution $\mathcal{N}(n/2,\frac{p}{1-p}\sigma_A^2(p=0.5))$ for the probability that 
the walker made $n_A$ of $n$ steps to the right \cite{Konno03}. Here $\mathcal{N}(X_0,\sigma^2)$ is the normal distribution 
with maximum at $X_0$ and variance $\sigma^2$, and $\sigma_A^2(p=0.5)=n/4$ is the variance for the number $n_A$ of 
steps to the right in the case of uncorrelated random walk ($p=0.5$). The position $x_n$ of the walker after $n$ steps is $n_A-n_B$.
Numerical calculations (see figure \ref{fig:numvstdl} in 
appendix \ref{appendix:figures}) show that for $n=100$ the thermodynamic limit is a 
good approximation (relative error for the standard deviation $<1\%$) for $0.2 < p < 0.8$ 
The probability $p$ to keep the direction should equal the sum of the correlation coefficients to find an 
event $A$ directly followed by an event $A$ and $B$ directly followed by $B$.
\begin{equation}
p \approx cc_{\text{keep direction}}=cc_k=P(\langle AA\rangle_{\Delta events=1})+P(\langle BB\rangle_{\Delta events=1})
\end{equation}
\par
We map the events of the oscillation experiment on a special correlated walk -- Gillis' 
random walk in one dimension \cite{Guillotin-Plantard05}:
\begin{itemize}
\item ``An agent of group $A$ passes the bottleneck." $\rightarrow$ ``Random walker moves one step to the right."
\item ``An agent of group $B$ passes the bottleneck." $\rightarrow$ ``Random walker moves one step to the left."
\end{itemize}
\par
Because of equation (\ref{eq:R}) there is a connection between $\langle R(n_A,n_B) \rangle$ - the expectation value 
of the number of runs - and $\sigma_A$. Since $\sigma_A=\sigma_B$ one has
\begin{equation}
\langle R(\sigma_A)\rangle  \approx \frac{2}{n} \Big(\frac{n}{2}+\sigma_A\Big)\Big(\frac{n}{2}-\sigma_A\Big) + 1 = \frac{n}{2}+1-2\frac{\sigma_A^2}{n}
\end{equation}

\section{Values for an uncorrelated process}
For $n=100$ and $N_A=N_B=202$
the expectation value for $n_A$ has to be 50 ($=n/2$) due to the symmetry of the scenario. Even if a finite group size effect
would in a strict sense imply a correlation, we here want to give not only the values for an uncorrelated process with no
finite group size effect, but also for an else uncorrelated process with finite group size effect. In the first case, where the passing
order can be imagined to be determined by a guardian at the bottleneck who decides about the right of passage of some member of one
of the two groups by throwing a coin, the basic distribution is binomial. (All numerical values are calculated for $n=100$ and $N_A=N_B=202$)
\begin{eqnarray}
P(n_A)&=&{n \choose n_A}2^{-n}\\
{\sigma_A}&=&\frac{\sqrt{n}}{2}=5\\
\overline{\langle R\rangle }&=&\sum_{n_A=0}^nP(n_A)\langle R(n_A,n_B) \rangle=\frac{n}{2}+\frac{1}{2}=50.5 \label {eq:Rexpav}
\end{eqnarray}
The latter case, where the guardian decides about the
right of passage with equal probability for each of the remaining individuals, is governed by the hypergeometric distribution.
\begin{eqnarray}
P(n_A)&=&\frac{{n \choose n_A}{N-n \choose N_A-n_A}}{{N \choose N_A}}\\
{\sigma_A}&=&\sqrt{\frac{n}{4}\bigg(\frac{N-n}{N-1}\bigg)}=\sqrt{\frac{7600}{403}}\approx 4.34\\
\overline{\langle R\rangle }&=&\frac{n}{2}+1-\frac{1}{2}\frac{N-n}{N-1}=50\frac{251}{403}\approx 50.62
\end{eqnarray}
It is assumed that the size of groups on both sides of the bottleneck has no influence on the movement probability as long as both
groups have some minimal size which should be exceeded by construction during the counting process as the largest possible inequality 
is 102:202. However even if this was not the case, this comparison shows that the expectation values of the
two possible cases of an uncorrelated process differ only slightly. We now can take these results as basis for comparison to the
results of our simulations.

\section{Oscillations without dynamic floor field}
\begin{figure}[htbp]
\begin{center}
\includegraphics[width=100pt]{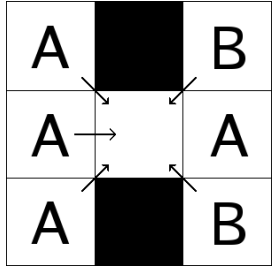}
\caption{One of twelve possible situations after an agent has passed.}
\label{fig:osc_passed}
\end{center}
\end{figure}
For $k_D=0$ there is some sort of ``intrinsic" correlation, which stems exclusively 
from the hard-core exclusion of the agents. As the last agent who passed, 
occupies one of the three cells which are adjacent to the bottleneck cell, there are on average 
fewer agents on the adjacent cells on the side of the group that could not move than on 
the other side. Figure \ref{fig:osc_passed} shows the situation when there is the maximum of
three agents of the group that moved last and the maximum of only two agents of the other 
group. Of course there can be other combinations (3:2, 3:1, 3:0, 2:2, 2:1, 2:0, 1:2, 1:1, 
1:0, 0:2, 0:1, 0:0) and the group that moved last can have fewer agents adjacent to the 
bottleneck, but on average there will be more. Therefore on that side there will on average 
also be more agents that plan to move to the bottleneck cell and so also the probability that 
one of them wins a possible conflict is increased, as the winner of a conflict is chosen with 
equal probability out of all competitiors. Note that this effect crucially depends on the choice of
von Neumann neighborhood or Moore neighborhood as neighborhood for $v_{max}=1$.\par
For $k_S=0.5$ we repeated
the calculations for the three cases $v_{max}=1$, $v_{max}=3$ and $v_{max}=5$ one-hundred times each. 
The tables of this and the following section show the average number of runs that occured,
the average number of runs the Wald-Wolfowitz test predicts for uncorrelated behavior, average z-values as well as 
the number of times (out of 100) the z-value was smaller than -1.95 (5\%-significance level)
and smaller than -2.58 (1\%-significance level), the standard deviation of $n_A$, the evacuation time (e.t.) in rounds and
the probability to continue in the same direction as calculated from $\sigma_A$ ($p$) as well as calculated from
correlation coefficients ($cc_k$).
\begin{center}
\begin{tabular}{c|ccccccccc}
$v$& $\bar{r}$ & $\overline{\langle R(n_A,n_B)\rangle }$ & $\bar{z}$ & 5\% & 1\% & $\sigma_A$ & $\overline{e.t.}$ & p($\sigma_A$)&$cc_k$\\ \hline
 1 &  37.04    &      50.51                &   -2.73   &  80 & 54  &    4.95    & 32945.9 & 0.49 & 0.64\\
 3 &  37.47    &      50.26                &   -2.61   &  76 & 53  &    6.05    & 25676.0 & 0.59 & 0.63\\ 
 5 &  37.94    &      50.36                &   -2.53   &  73 & 51  &    5.60    & 21395.4 & 0.56 & 0.63\\
\end{tabular}
\end{center}
Note that $\overline{\langle R(n_A,n_B)\rangle }$ is the average of the values for $\langle R(n_A,n_B)\rangle$ calculated from the 100 simulation
results for $n_A$ respectively $n_B$ using equation (\ref{eq:R}) and averaged in the way it is done in equation (\ref{eq:Rexpav}) however using the distribution of simulation results and not a theoretical distribution $P(n_A)$. The other averages are also averages over 100 simulations.

\section{Oscillations with dynamic floor field}
For all simulations $k_D=0.3$ and $k_S=0.5$ has been set. This is quite a small value, however for $k_S=1.0$ or even larger the crowds
on both sides of the bottleneck become too dense, such that it becomes difficult for an agent to pass through them after he has 
passed the bottleneck. For each set of parameters the simulation was repeated 100 times.
We mainly experimented with the maximum speed $v_{max}$ (which was equal for all agents in a scenario) 
and the parameters for diffusion $\alpha$ and decay $\delta$. 
The largest deviations from the uncorrelated case are marked in the following tables.
\setlength{\tabcolsep}{4pt}
\begin{center}
\begin{tabular}{c|ccccccccc}
\multicolumn{10}{c}{\boldmath $v_{max}=1$, \boldmath $\delta=0.03$} \\
$\alpha$&$\bar{r}$&$\overline{\langle R(n_A,n_B)\rangle }$&$\bar{z}$& 5\% & 1\% &$\sigma_A$&$\overline{e.t.}$&p($\sigma_A$)&$cc_k$\\ \hline
0.03    &  25.87  &             47.70                     &{\bf -4.69}&{\bf98}&{\bf95}&12.84&      5765.14   &     0.87    & 0.75 \\
0.10    &  28.05  &             47.15                     &     -4.15 &    94 &    84 &13.84&      6642.97   &     0.88    & 0.73 \\
0.30    &  32.87  &             50.08                     &     -3.52 &    93 &    86 & 6.75&     13479.4    &     0.65    & 0.68 \\ 
1.00    &  34.06  &             49.93                     &     -3.26 &    90 &    75 & 7.32&     13375.8    &     0.68    & 0.67 \\
\end{tabular}
\end{center}
\begin{center}
\begin{tabular}{c|ccccccccc}
\multicolumn{10}{c}{\boldmath $v_{max}=3$, \boldmath $\delta=0.03$} \\
$\alpha$&$\bar{r}$&$\overline{\langle R(n_A,n_B)\rangle }$&$\bar{z}$& 5\% & 1\% &$\sigma_A$&$\overline{e.t.}$&p($\sigma_A$)&$cc_k$\\ \hline
0.03    &{\bf 15.54}&            27.68                    &   -3.86 &  74 &  69 &      34.06 &     2277.85   &     0.98    & 0.85 \\
0.10    &     19.59 &            33.63                    &   -4.09 &  81 &  73 &      29.47 &     2058.30   &     0.97    & 0.81 \\
0.30    &     15.75 &       {\bf 25.22}                   &   -3.61 &  74 &  65 & {\bf 35.79}&     1821.15   &     0.98    & 0.85 \\ 
1.00    &     29.16 &            40.47                    &   -2.86 &  74 &  58 &      22.94 &     2214.92   &     0.95    & 0.72 \\
\end{tabular} 
\end{center}
\begin{center}
\begin{tabular}{c|ccccccccc}
\multicolumn{10}{c}{\boldmath $v_{max}=5$, \boldmath $\delta=0.03$} \\
$\alpha$&$\bar{r}$&$\overline{\langle R(n_A,n_B)\rangle }$&$\bar{z}$& 5\% & 1\% &$\sigma_A$&$\overline{e.t.}$&p($\sigma_A$)&$cc_k$\\ \hline
0.03    & 21.60   &              25.86                    &   -1.65 &  40 &  28 &    35.34 &       2063.88   &     0.98    & 0.79 \\
0.10    & 20.85   &              31.29                    &   -3.09 &  61 &  55 &    31.34 &       1758.01   &     0.98    & 0.80 \\
0.30    & 17.68   &              27.58                    &   -3.29 &  64 &  57 &    34.18 &  {\bf 1519.78}  &     0.98    & 0.83 \\ 
1.00    & 28.22   &              36.51                    &   -2.41 &  57 &  46 &    26.87 &       1730.61   &     0.97    & 0.73 \\
\end{tabular} 
\end{center}
(Also see figures \ref{fig:zvonalpha}, \ref{fig:etvonalpha}, \ref{fig:sigmavonalpha}, \ref{fig:cor_v1}, 
\ref{fig:cor_v3} and \ref{fig:cor_v5} in appendix \ref{appendix:figures}.)
Other simulations showed that for $\delta>0.1$ the effect begins to vanish as the dynamic floor field decays 
too fast to have a significant influence and for $\delta<0.01$ the effect is hidden by the dominant dynamic 
floor field that then induces irrational behavior. The agents begin to move in circles instead of heading to 
the bottleneck. Note that effects like this can also occur for too small $\alpha$, if the dynamic floor field
becomes too strong on some cells.\par
Compared to the case without dynamic floor field, the dynamic floor field with its time dependence brings in an another
dimension that can be analyzed: As long as the dynamic floor field has not reached a steady-state, the 
oscillation will be time dependent. This implies that the measured values either change with $n$ or if $n$ is
kept constant but the measurement process starts not with the first event but later. For $\alpha=1.0$, $v_{max}=1$ and $n=150$ 
we got $z=-3.92$, which is quite a difference to $z=-3.26$ for $n=100$. However except for the case $\alpha=0$, where they
became slightly larger, the correlation coefficients did not change at all. As the existence of time-dependent
oscillation types is model-dependent, the analysis of this phenomenon shall not be done in further detail in this work.

\section{Discussion}
Some observations made in the $k_D=0.3$ data are as follows.
\begin{itemize}
\item The evacuation time (in rounds) drops dramatically compared to $k_D=0$ simulations. So it can be assumed that the larger groups of agents passing the door represent a more efficient behavior. This can be interpreted as less time being consumed by conflict solution processes at the bottleneck.\vspace{-6pt}
\item The minimal $z$-value for each $v_{max}$ is found at larger $\alpha$ for larger $v_{max}$. (See also figure \ref{fig:zvonalpha}.)\vspace{-6pt}
\item The same holds for the minimum of the evacuation time. (See also figure \ref{fig:etvonalpha}.)\vspace{-6pt}
\item Correlations manifest themselves for $v_{max}=1$ typically in smaller $z$-values and for $v_{max}=5$ typically in larger $\sigma_A$.\vspace{-6pt}
\item $\sigma_A$ for $v_{max}=3$ and $v_{max}=5$ are that large that $(n_A,n_B>10)$ is not always fulfilled. 
\item For simulations with dynamic floor field, the correlations are stronger for $v_{max}=3$ and $v_{max}=5$ than for $v_{max}=1$. This is due to the larger area of influence for higher speeds: more agents could choose the bottleneck cell as their destination cell. But due to the direction of the dynamic floor field mainly agents of the group that moved last indeed do choose the bottleneck cell as destination cell, which leads to a stronger outnumbering of agents attempting to follow one of their group on the bottleneck cell compared to agents trying to change the direction of the flow, than in the case of $v_{max}=1$.
\item Also for simulations with dynamic floor field $p(\sigma_A)$ is most of the time much larger than $cc_k$, however within each $v_{max}$-set of results there appears to be a tendency that $cc_k$ and $p$ are positively correlated. The difference points to a dynamic floor field that is too strong to reverse direction with only one agent passing. If for example after a sequence of agents of group $A$ one agent of group $B$ passes the bottleneck, a typical sequence will look like $AAAAAABAAAA$ if the dynamic floor field does not change direction with that single agent. If however it does change, a typical sequence would be       $AAAAAABBBBB$. In the first case $\sigma_A$ will grow, as the dominance of agents of one group outlasts the accidental event with small probability that one agent of the other group passes. In the latter case $\sigma_A$ will be comparatively small since one long run of $A$s can be followed by an equally long run of $B$s and vice versa. The correlation coefficient $cc_k$ however is much less affected by this phenomenon as it is not distinguished between $AA$ and $BB$ sequences. Take for illustration an $A$-$B$-symmetric $n=11$ example: ($AAAAAABAAAA$, $BBBBBBABBBB$) has $\sigma_A=4.5$ and $cc_k=0.8$ and ($AAAAAABBBBB$, $BBBBBBAAAAA$) has $\sigma_A=0.5$ and $cc_k=0.9$. So while $\sigma_A$ becomes larger with a stronger dynamic floor field, $cc_k$ becomes larger. Consequently did a simulation with $v=3$, $\alpha=1$ and $\delta=0.1$ result in $cc_k=0.64$ and $p=0.61$ instead of $cc_k=0.72$ and $p=0.95$ as for the $\delta=0.03$ simulation. This shows that a comparison of $p$ and $cc_k$ can give a hint that $\delta$ was chosen too small, as for real pedestrians the passing of one individual is enough to completely change the odds.
\item Simulations without dynamic floor field gave values for $\sigma_A$ and $p$ that might make $p$ look dependent of $v$. However the reason for this is a relatively broad distribution of the $\sigma_A$. In 20 additional $v_{max}=1$ calculations of $\sigma_A$ with $n=100$ simulation repetitions each $\sigma_A$ variied between $\sigma_A=5.34$ and $\sigma_A=6.72$ this implies values for $p$ between $p=0.53$ and $p=0.64$ with an average of $p=0.575$. At the same time $cc_k$ remained relatively constant between $cc_k=0.62$ and $cc_k=0.64$. That $p$ even at average remains smaller than $cc_k$ probably results from a reduced local density directly in front of the bottleneck if a group had a sequence of agents passing the bottleneck. The reduced local density then increases the probability of a change of the flow direction. This is an effect of the kind of a hypergeometric distribution, that reduces the standard deviation (see above). Here too $p$ is much more affected than $cc_k$ as $cc_k$ is not affected by the relative total numbers of $A$s and $B$s within the first $n$ events, but only by the number of changes of the flow direction. 
\end{itemize}
The significant drop in the evacuation time compared to $k_D=0$ is a sign that the dynamic floor field not only leads to a more efficient
motion into the bottleneck but also out of the bottleneck and through the group on the other side.\par
It is not very surprising, that the effects of the dynamic floor field are strongest for larger $v_{max}$ at larger $\alpha$, since
the larger neighborhood in which faster agents can move during one round makes it necessary that also the dynamic floor field diffuses
faster to keep up with the agents.\par
If $\sigma_A$ becomes too large, moderate $z$-values are no indicator that the events are uncorrelated. Contrary to that
the scenario is just too small for such sets of parameters to make large $z$-values possible.\par

\section{Summary, conclusions and outlook}
A model of pedestrian motion was presented, which is able to modulate the characteristic observables of oscillations.
The model was tested concerning this capability using the Wald-Wolfowitz test, a comparison to correlated random walks and a calculation of
correlation coefficients.
This framework for characterizing should be applicable to any model of pedestrian motion as well as corresponding experimental 
and empirical data. However these observables would show details that are far from
having been measured in reality at present. Anyway the scenario of oscillations at bottlenecks did show up a rich variety of phenomena and observables, which might be used to characterize and distinguish models of pedestrian motion. The investigation of time-dependent oscillation types as well as the distribution of the length of runs could be further elements in such discussions. Observations and experiments in the field would be of great interest.\\
Since our model is able to reproduce quite different types of oscillations it probably will also be able to 
reproduce the range of oscillations that occur in reality.

\section{Acknowledgments}
This work has been financed by the ``Bundesministerium f\"ur Bildung und Forschung" (BMBF) within the PeSOS project. We would like
to thank A. Schadschneider for comments and discussion.

\nocite{PED01,PED05}
\bibliographystyle{unsrt}
\bibliography{referenzen}

\begin{appendix}
\section{Figures} \label{appendix:figures}

\begin{figure}[htbp]
\begin{center}
\includegraphics[width=300pt]{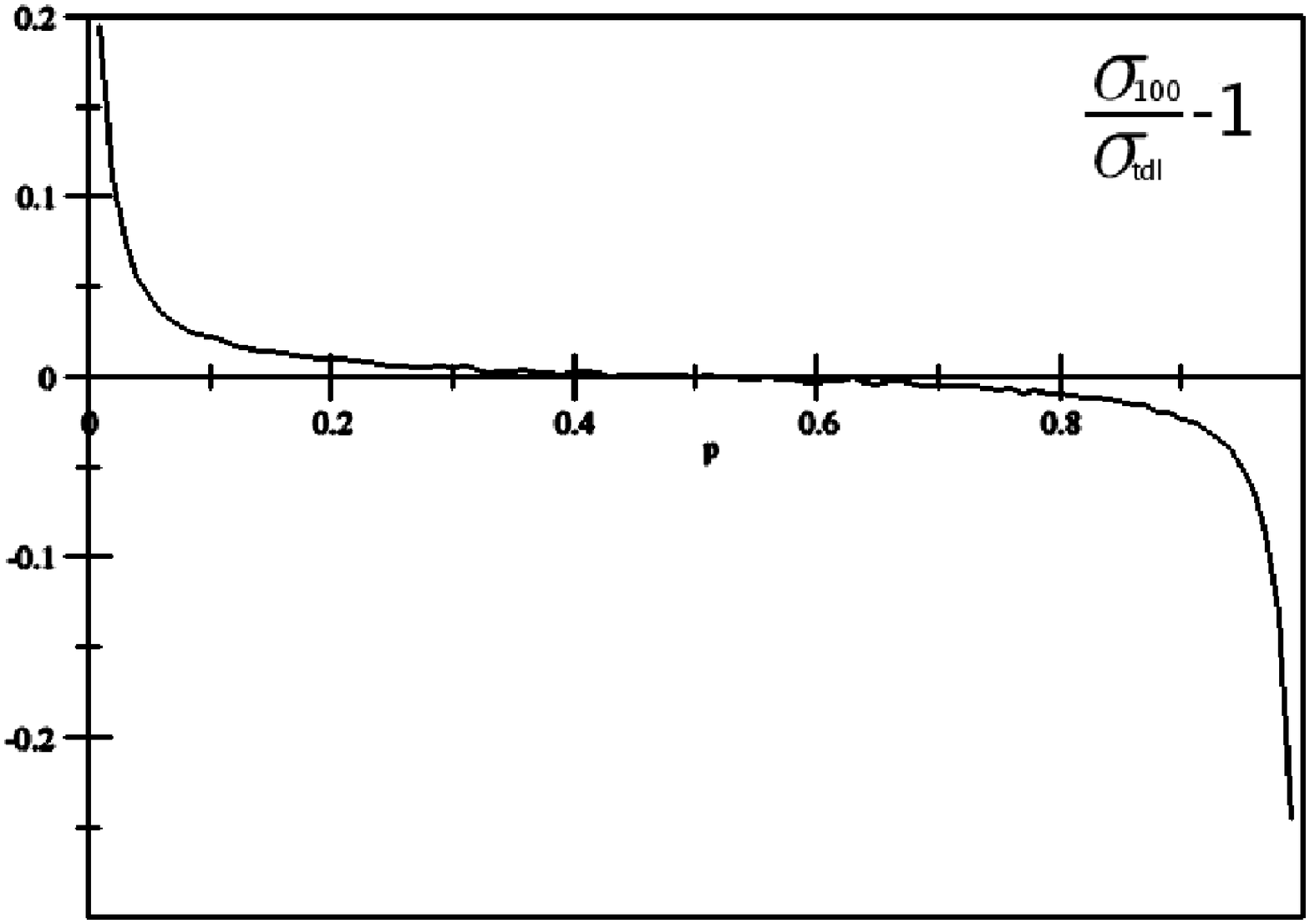}
\caption{Relative difference of the numerically calculated standard deviation for the number of steps into a certain direction for $n=100$ steps against the theoretical thermodynamical ($n\rightarrow \infty)$ limit of the same standard deviation in dependence of the probability $p$ to continue motion in the same direction.}
\label{fig:numvstdl}
\end{center}
\end{figure}

\begin{figure}[htbp]
\begin{center}
\includegraphics[width=300pt]{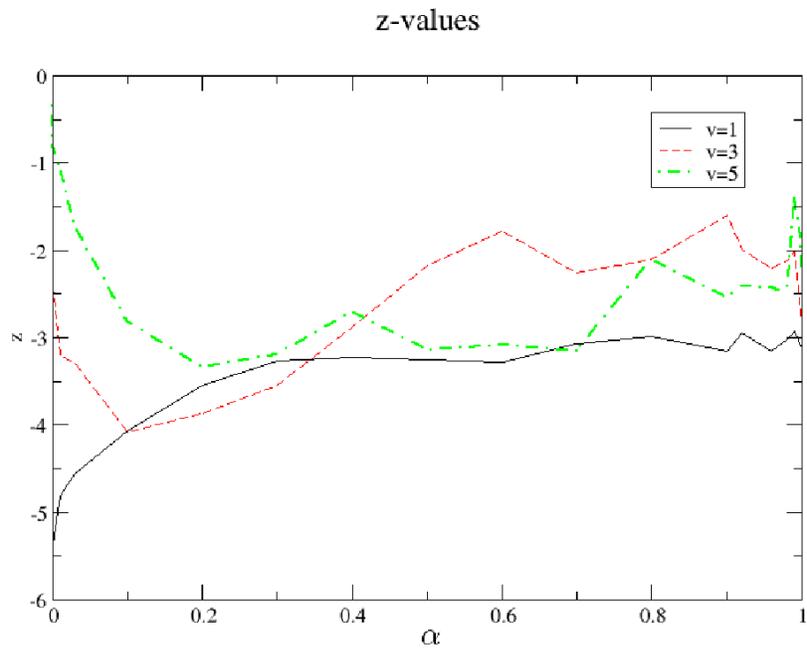}
\caption{Average $z$-value for different $\alpha$ at $\delta=0.03$.}
\label{fig:zvonalpha}
\end{center}
\end{figure}

\begin{figure}[htbp]
\begin{center}
\includegraphics[width=300pt]{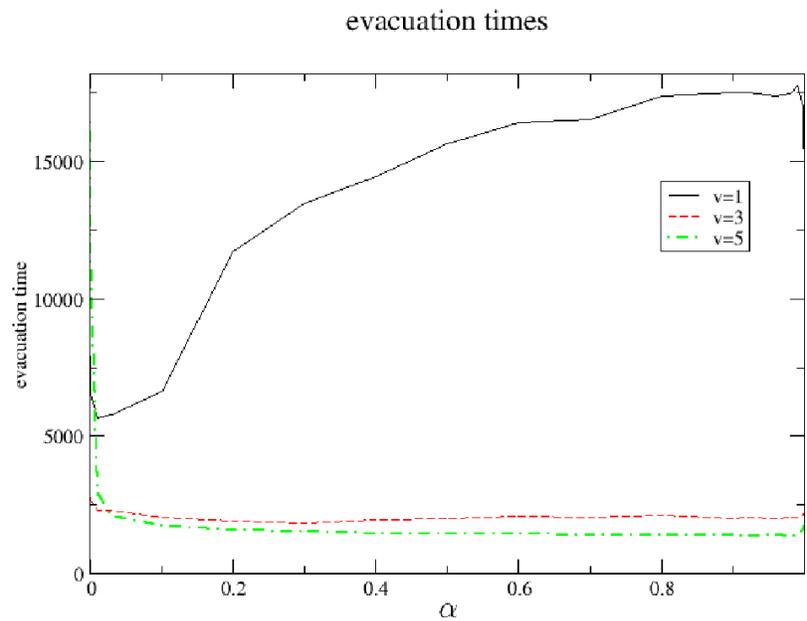}
\caption{Average evacuation times for different $\alpha$ at $\delta=0.03$.}
\label{fig:etvonalpha}
\end{center}
\end{figure}

\begin{figure}[htbp]
\begin{center}
\includegraphics[width=300pt]{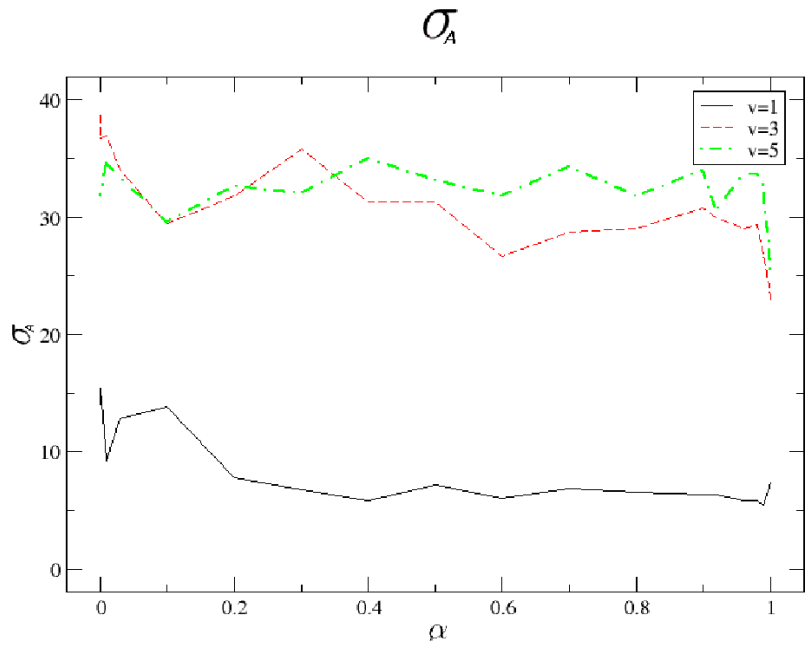}
\caption{$\sigma_A$ for different $\alpha$ at $\delta=0.03$.}
\label{fig:sigmavonalpha}
\end{center}
\end{figure}

\begin{figure}[htbp]
\begin{center}
\includegraphics[width=300pt]{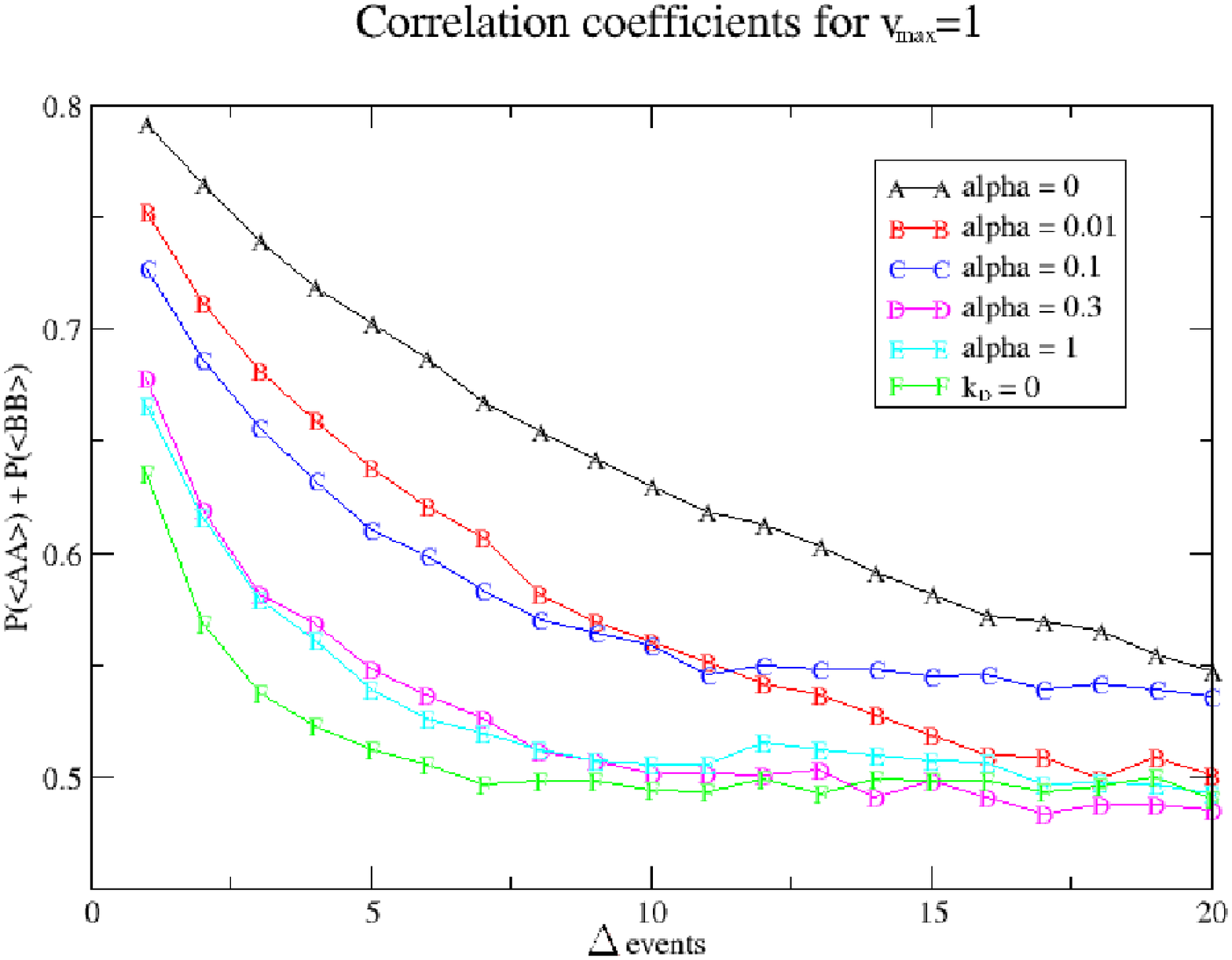}
\caption{Correlation coefficients for different $\alpha$ and event distance at $v_{max}=1$}
\label{fig:cor_v1}
\end{center}
\end{figure}

\begin{figure}[htbp]
\begin{center}
\includegraphics[width=300pt]{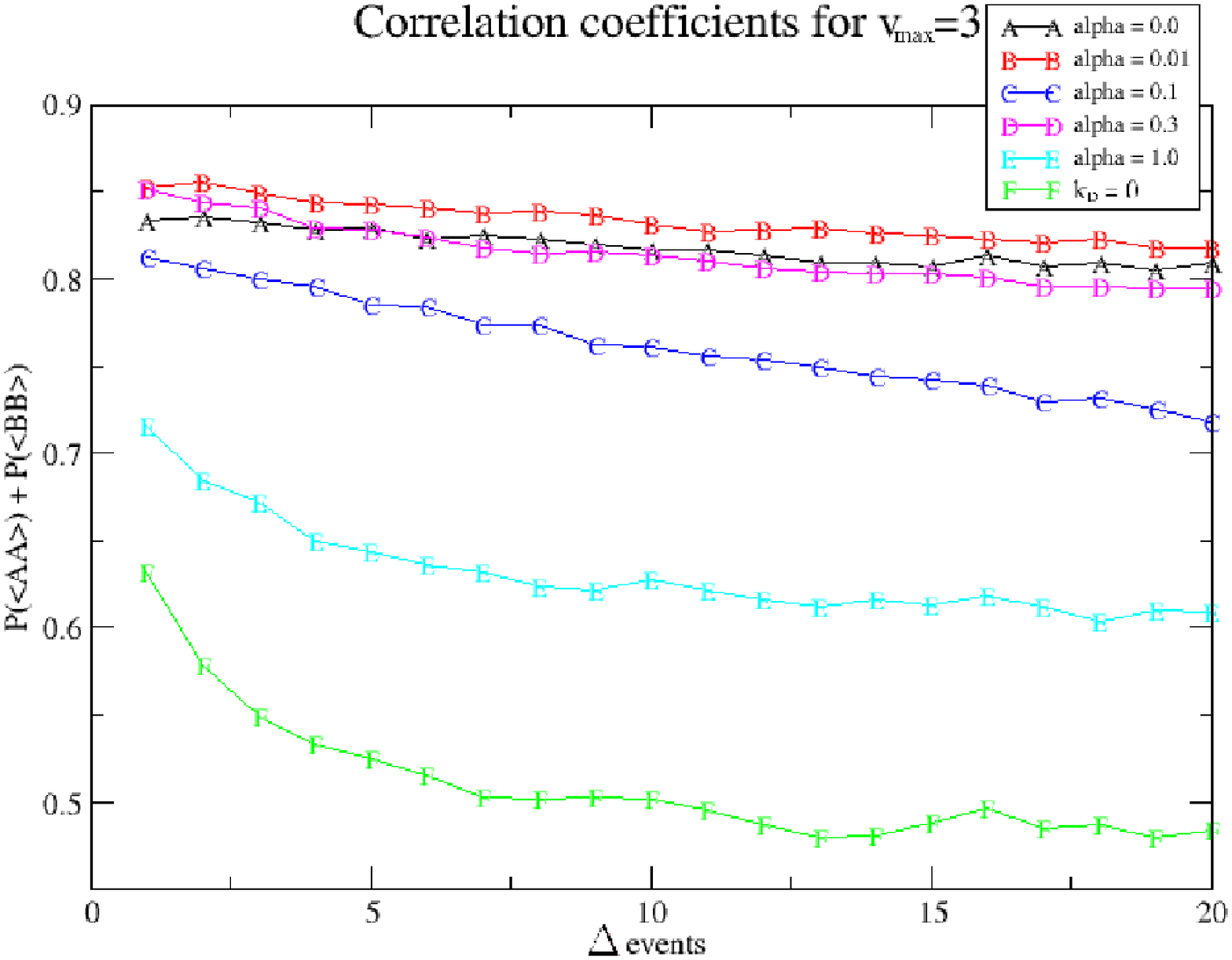}
\caption{Correlation coefficients for different $\alpha$ and event distance at $v_{max}=3$}
\label{fig:cor_v3}
\end{center}
\end{figure}

\begin{figure}[htbp]
\begin{center}
\includegraphics[width=300pt]{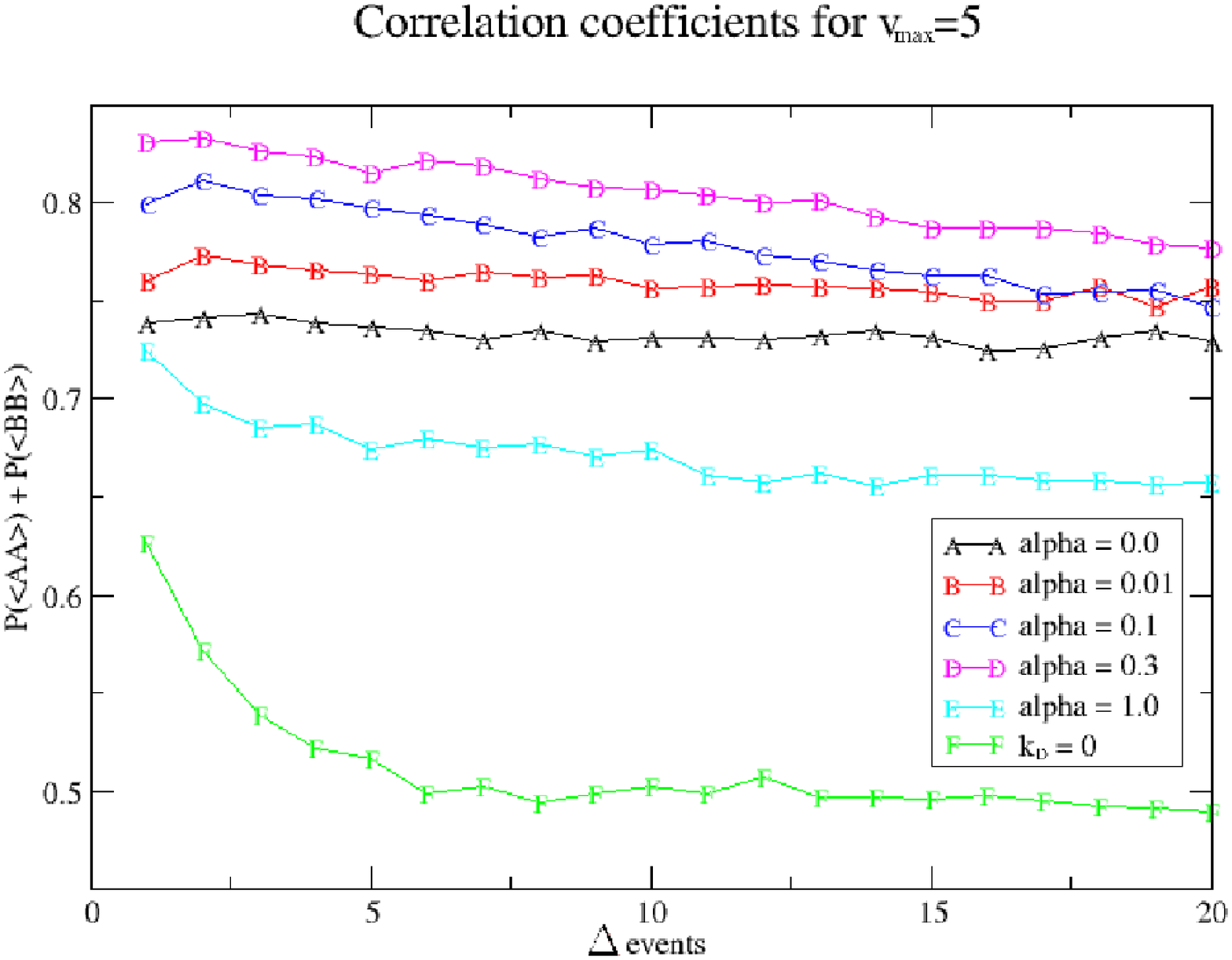}
\caption{Correlation coefficients for different $\alpha$ and event distance at $v_{max}=5$.}
\label{fig:cor_v5}
\end{center}
\end{figure}

\end{appendix}

\end{document}